\documentstyle[ltwol, epsfig]{article}

\def\be{\begin{equation}}
\def\ee{\end{equation}}
\def\bea{\begin{eqnarray}}
\def\eea{\end{eqnarray}}

\begin{document}

\title{{\sl BLAST} -- A {\bf B}alloon-borne {\bf L}arge {\bf A}perture 
{\bf S}ubmillimeter {\bf T}elescope}

\author{Mark J. Devlin for the {\sl BLAST} Collaboration \footnotemark[1]}

\address{Department of Physics and Astronomy, University of Pennsylvania, 
Philadelphia, PA, 19104, USA\\E-mail: devlin@physics.upenn.edu}

\twocolumn[\maketitle\abstracts{A new generation of sub-orbital
platforms  will be operational in the next few years.  These new
telescopes  will operate from airborne and balloon-borne platforms
where the atmosphere is transparent enough to allow sensitive
measurements to be made in the submillimeter bands. The telescopes will
take advantage of state-of-the-art instrumentation including large
format bolometer arrays and spectrometers. Other papers in this volume
will deal specifically with the potential of these bands. In this
paper will review
the capabilities the  BLAST balloon-borne telescope.}]

\section{Introduction to {\sl BLAST}}
\vspace{-.1in}
\footnotetext[1]{The {\sl BLAST} collaboration includes:\\
Peter Ade - Queen Mary and Westfield College, London, UK\\
James Bock - JPL, Pasedena, CA, USA\\
Paolo DeBernardis - University of Rome, Rome, IT\\
Mark Devlin - University of Pennsylvania, Philadelphia, PA, USA\\
Joshua Gundersen - Princeton University, Princeton, NJ, USA\\
Mark Halpern - University of British Columbia, Vancouver, CA\\
David Hughes - INAOE, Puebla, MX\\
Jeff Klein - University of Pennsylvania, Philadelphia, PA\\
Phillip Mauskopf - Cardiff University, Cardiff, UK\\
Silvia Masi - University of Rome, Rome, Italy\\
Barth Netterfield - University of Toronto - Toronto, Canada\\
Luca Olmi - University of Massachusetts, Amherst, MA, USA\\
Lyman Page - Princeton University, Princeton, NJ, USA\\
Douglas Scott - University of British Columbia, Vancouver, Canada\\
Gregory Tucker - Brown University, Providence, RI, USA }

The ``Balloon-borne
Large-Aperture Sub-millimeter Telescope'' ({\sl BLAST}) incorporates a
2.0~m mirror with 
plans to increase to a 2.5~m mirror.  The telescope will operate on a
Long  Duration Balloon 
(LDB) platform with large format bolometer arrays at 250, 350 and
500~$\mu$m.
{\sl BLAST} will address some of the most important
Galactic and cosmological questions regarding the formation and
evolution of stars, galaxies and clusters.   It will conduct large-area
sensitive Galactic and
extragalactic surveys which will: (i) identify large numbers of distant
high-redshift galaxies;
(ii) measure cold pre-stellar sources associated with the earliest
stages of star and planet formation; (iii) make 
high-resolution  maps of diffuse Galactic emission from low to high
Galactic latitudes.

The primary advantage of {\sl BLAST} over existing and
planned
bolometer arrays such as SCUBA\cite{hol99} on the JCMT, SHARC \cite{cso}
on the CSO (including their respective upgrades) 
and HAWC on SOFIA \cite{harper} is the dramatically
increased atmospheric transmission at balloon altitudes
which results in greatly enhanced sensitivity at wavelengths 
$\le 500\,\mu$m.  {\sl BLAST} complements FIRST satellite by overlapping
the FIRST frequency coverage and will have the ability
to test new technologies for future space-based missions. 
{\sl BLAST} is the only instrument with sufficient field of view,
sensitivity, and integration time to conduct follow-up surveys of SIRTF/MIPS
at 200-400~$\mu$m.

\section{Science Goals of {\sl BLAST}}

{\sl BLAST} will be the first balloon-borne telescope
to take advantage of the bolometric focal plane arrays being developed
for {\sl FIRST}.   It will be capable of probing the sub-mm 
with high spatial resolution and sensitivity providing the opportunity
to conduct unique Galactic and extragalactic surveys. 
Compared to the pioneering flights
of PRONAOS \cite{lam98}, {\sl BLAST} will have an advantage of $>100$
times the mapping speed.
The scientific motivations for {\sl BLAST} are similar to those of {\sl
FIRST} and
are achievable within 3--5 years with a series of LDB flights. 

We expect {\sl BLAST} to achieve the following science goals:

\begin{itemize}

\item Conduct large-area extragalactic 200--500~$\mu$m surveys 
and detect $\sim 150$ and $\sim 1500$ high-$z$ galaxies 
in the test flight and long-duration flights respectively. 

\item Measure the confusion noise at 200--500~$\mu$m 
thereby laying the foundation for future {\em BLAST} and {\sl FIRST} survey
strategies.  It will also allow a study of the clustering of
dust-emitting galaxies over an important range of angular scales.

\item Combine the {\sl BLAST} 200--500~$\mu$m spectral energy
distributions (SEDs) and source counts
with those of SCUBA at 850~$\mu$m. 
This will determine the redshifts, rest-frame luminosities, 
star formation rates (SFRs),  and evolutionary history of starburst
galaxies in the high-$z$ universe   It will aslo identify the galaxy
populations  
responsible for producing the far-IR background.  

\item Conduct Galactic surveys of molecular clouds and 
identify dense, cold pre-stellar (Class--0) cores associated with the
earliest stages of star formation.  Combining the 200-500~$\mu$m {\sl BLAST}
data with SCUBA data at 850~$\mu$m will determine their density and temperature
structures which are sensitive to the details of the clouds collapse.

\item Survey the diffuse Galactic emission  and make detailed
comparisons with surveys at longer and shorter
wavelengths (CGPS, SCUBA, MSX).

\item Observe objects within our solar system including the planets
and large asteroids. 
 
\end{itemize}

\section{Extragalactic Surveys}
\begin{figure}[tbh]
\centerline{
\psfig{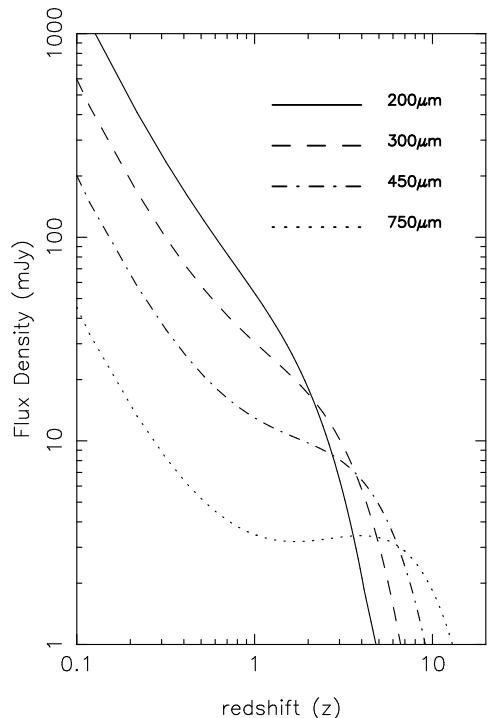}}
\vspace{-0.15in}
\caption{Flux density vs. redshift for luminous 
($L_{\rm{FIR}}/L_{\odot} \sim 10^{12}$) starburst galaxies.}
\label{fig:fluxz}
\end{figure}

The first extragalactic sub-mm (850 micron) surveys
have already been completed by SCUBA covering areas of 0.002--0.12~deg$^2$ with
respective $3\sigma$ depths in the range $ 1.5 < S(\rm{mJy}) < 8$
 \cite{smail97,dhh98,lilly99,cha00}. Observations of
starburst galaxies in the high-$z$ universe at sub-mm wavelengths have a
particular advantage compared to the optical and FIR due to the strong
negative k-correction which enhances the observed sub-mm fluxes by
factors of 3--10 at $z > 1$ (Figure \ref{fig:fluxz}).  The
combination of SCUBA surveys with the necessary shorter wavelength
sub-mm data from experiments like {\sl BLAST} (and later from {\sl FIRST})
will provide powerful constraints on our understanding of the models and
processes by which galaxies and clusters form and subsequently evolve.
The following preliminary results from SCUBA surveys (see other
contributions to this Proceedings) alone have already
made a significant impact on several cosmological questions and have
demonstrated the importance of further shorter wavelength sub-mm
cosmological studies:

\begin{itemize}

\item
Approximately 30--50\% of the FIR--sub-mm background detected by COBE
has been resolved into individual galaxies at flux densities 
$ S_{850\mu \rm{m}}>2\,$~mJy.  Existing surveys,
which are confusion-limited at about this flux level,
are within a factor of only a few in
sensitivity of resolving the entire sub-mm/FIR background.

\item 
Sub-mm sources generally appear to be associated with
$z > 1$ optical and weak radio galaxies, although there is still much
debate about the fraction of sources at $z\geq 2$.

\item
The faint sub-mm source-counts at 850~$\mu$m are reasonably well
determined between 1--8\,mJy and significantly exceed a no-evolution
model, requiring roughly $(1+z)^{3}$ evolution out to $z\sim1$, but with
little constraint at higher redshifts.

\item
At high-redshift ($ 2 < z < 4$) the sub-mm surveys
appear to find $\sim5$ times the star formation rate observed in the
optical surveys, although the effects of
dust obscuration and incompleteness in the optical are still debated.
\end{itemize}

\subsection{Why an LDB experiment at 200--500~$\mu$m?}


Detailed interpretation of the sub-mm sources is severely hindered by
uncertainties in their redshift distributions and luminosities.
These
uncertainties result directly from the sub-mm positional errors of
$\sim 3^{\prime\prime}$ that are typical for even the highest S/N
sub-mm detections, and from the lack of sub-mm data measuring the
redshifted FIR spectral peak at 200--450~$\mu$m \cite{dhh98}.

\begin{figure}[tbh]
\centerline{
\psfig{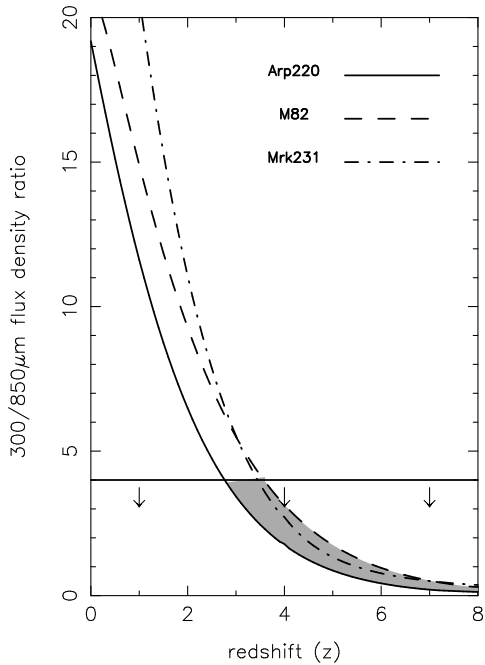}}
\vspace{-0.15in}
\caption {300/850~$\mu$m flux ratio vs. redshift 
for Arp220, Mrk231 and M82.
The shaded area shows the redshift range  consistent
with SCUBA sources $>13$~mJy with no {\sl BLAST} 5$\sigma$ 
counterparts at 300~$\mu$m.}
\label{fig:ratio}
\end{figure}

This ambiguity in the redshift of individual sub-mm sources will be
resolved with the {\sl BLAST} surveys. 
The complementary depths of the {\sl BLAST}
200-500~$\mu$m and SCUBA 850~$\mu$m surveys provide a measure of the
200/850~$\mu$m flux ratio, which is an extremely powerful and independent
diagnostic of the redshift (Figure \ref{fig:ratio}).
For example if there are 5$\sigma$ SCUBA sources ($S_{850\,\mu \rm{m}} >
13$~mJy) with no {\sl BLAST} 5$\sigma$ counterparts at 300~$\mu$m, {\it
i.e.} $S_{300\,\mu \rm{m}} < 50$~mJy, then the 300/850~$\mu$m flux
ratio must be $\leq 4$.  This implies that the sub-mm source must be a
galaxy at $z \geq 3$ for all typical starburst SEDs.
 
\begin{figure}[tbh]
\centerline{
\psfig{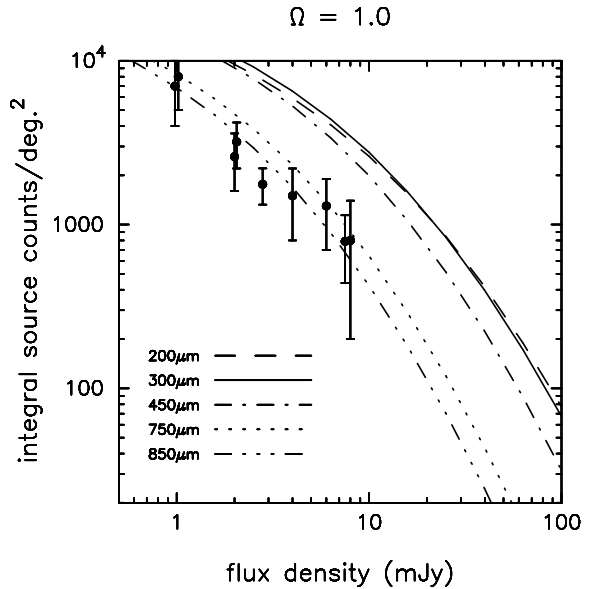}}
\vspace{-0.15in}
\caption{Predicted {\sl BLAST} source counts.
The shorter wavelength source counts are extrapolations of the
best-fit model at 850~$\mu$m. }
\label{fig:counts}
\end{figure}

\begin{figure}[tbh]
\centerline{
\psfig{file=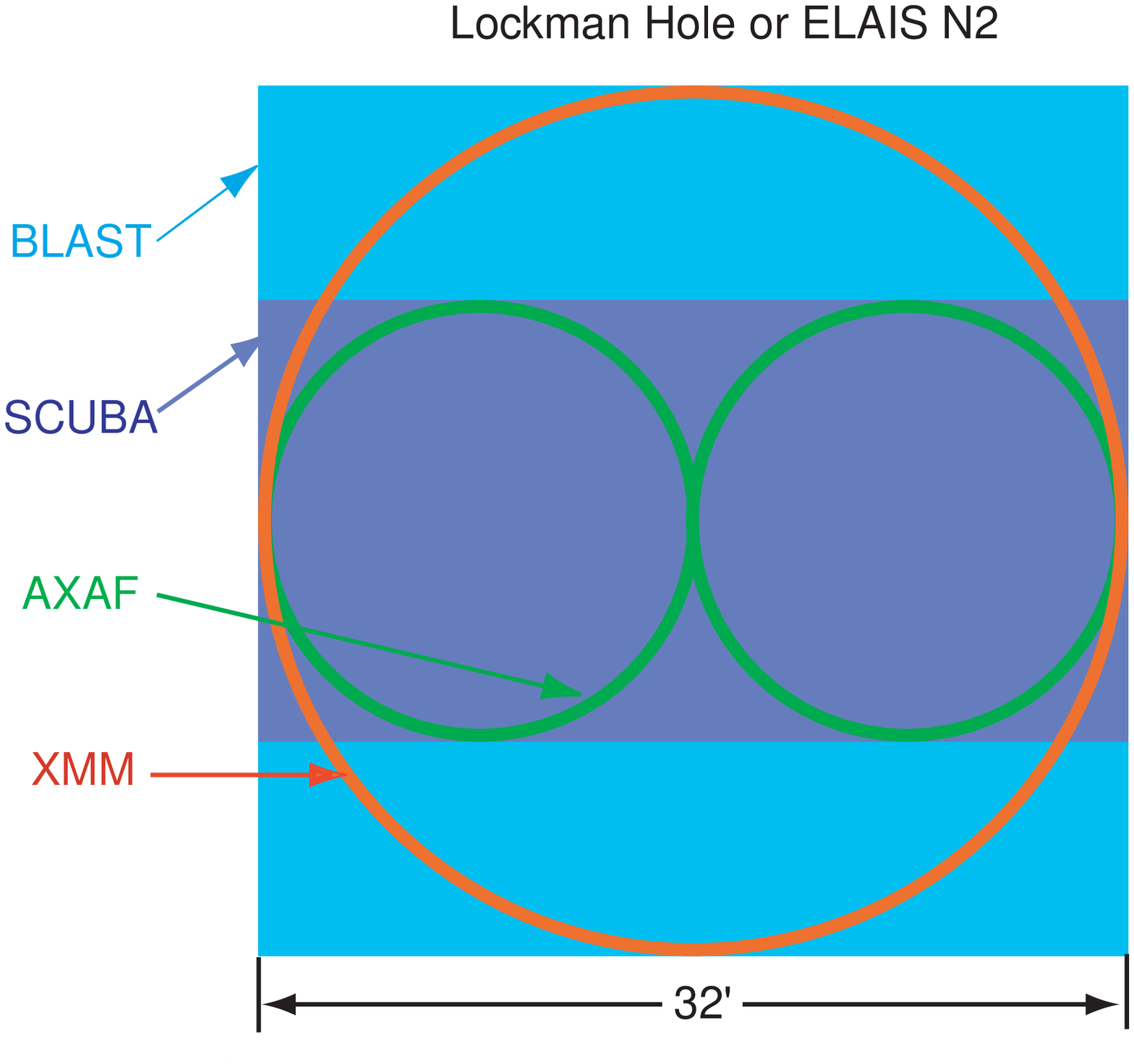,angle=0,width=2.75in}}
\vspace{-0.25in}
\caption{Coverage for a 6 hour, 0.55~deg$^2$ survey. 
The {\sl BLAST} survey is designed to match the coverage of the Lockman
Hole and ELAIS N2 fields at other wavelengths.}
\label{fig:survey}
\end{figure}

\subsection{Predicted Source-counts and Extragalactic Confusion}

The {\sl BLAST} surveys will measure the surface-density of the brighter
sources (20--1000\,mJy) at $200-500\,\mu$m, extending the flux range of
the sub-mm source-counts from $\sim 15$ to $\sim 1000$.  This will
allow an accurate determination of the model that describes the
evolution of high-$z$ starburst galaxies \cite{guider98,blain99}.  A
prediction of the 200--$500\,\mu$m
source-counts can be derived from preliminary evolutionary models that
fit the observed counts at FIR and sub-mm-wavelengths (Figure
\ref{fig:counts}).

Due to extragalactic sources, SCUBA 850~$\mu$m surveys with 15$''$
resolution are confused at a depth of $\rm 3\sigma < 2\,mJy$ which
corresponds to a source density $\rm \sim 4000 \pm 1500\, deg^{-2}$
 \cite{dhh98}.  Extrapolation to shorter wavelengths implies that
confusion at $300\,\mu$m will begin to become significant at
$3\sigma \sim 25$--$35\,$mJy. Representative estimates of the
extragalactic 3$\sigma$ confusion noise at 300--500~$\mu$m are 30\,mJy
for a primary mirror diameter of 2.0~m
(appropriate for the test-flight) and 25\,mJy for a 2.5~m primary
(planned for the LDB flight).  The initial {\sl BLAST} surveys 
will measure the true confusion level and hence will
influence the strategies for both future {\sl BLAST} and FIRST surveys
at 200--500~$\mu$m.  A statistical $P(D)$ analysis 
and correlation analysis \cite{borIP} of the noise in our
surveys will yield information on the source counts well below the
conventional confusion limit.

During the first 6 hour test flight {\sl BLAST} may
conduct a $0.55\,{\rm deg}^2$ extragalactic survey of the Lockman Hole
and ISO ELAIS regions at 300--750~$\mu$m with a 1$\sigma$ sensitivity of
$\sim 10$~mJy.  The alternative surveys outlined in
table~3.1 will detect similar numbers of galaxies at the
different flux levels, and therefore we can choose to match the {\sl
BLAST} surveys to the coverage of the existing and future 850~$\mu$m
SCUBA and 1100~$\mu$m BOLOCAM/CSO surveys in addition to the future
high-resolution deep AXAF and XMM surveys of these regions (Figure
\ref{fig:survey}).  The
X-ray follow-up data are crucial for determining the fraction of AGN in
such samples \cite{almIP}.

By combining these data with mid-IR observations from SIRTF
we can accurately measure the spectral energy distributions across the
rest-frame FIR peak, constrain the redshifts, bolometric luminosities,
SFRs and evolutionary history of high-$z$ starburst galaxies.  Together
with a deep 20~cm VLA survey \cite{ciliegi}, 
deep near-IR and optical imaging, we will
determine the nature of the population of high-$z$ galaxies that
contribute the dominant fraction of the extragalactic FIR-mm
background.\cite{lag,puget,sco00}

\begin{table*}[tbh]
\vspace{0.2cm}
\begin{center}
\begin{tabular}{ccccccc}
\multicolumn{7}{c}{6 hour 300~$\mu$m test-flight survey strategies: D=2.0~m,
$\theta = 41''$ } \\  
\hline \hline
survey area         & 
1$\sigma$ depth     & 
no. of pixels       & 
\multicolumn{2}{c}{no. of detected galaxies}   & 
\multicolumn{2}{c}{no. of $>5\sigma$ galaxies} \\ 

(sq. degrees) & 
              & 
              & 
$> 5  \sigma$  & 
$> 10 \sigma$  & 
$ z > 1$   & 
$ z > 3$   \\
\hline
0.24 & 7  mJy & 2352  & 120 & 34 & 110 & 18 \\
0.55 & 10 mJy & 4800  & 150 & 40 & 135 & 20 \\
1.1  & 15 mJy & 10800 & 135 & 30 & 125 & 16 \\
4.4  & 30 mJy & 43200 & 120 & 30 & 110 & 13 \\
\hline
\end{tabular}
\end{center}
\begin{center}
\begin{tabular}{ccccccc}
\multicolumn{7}{c}{50 hour 300~$\mu$m LDB survey strategies: D=2.5~m,
$\theta = 32''$  } \\  
\hline \hline
survey area         & 
1$\sigma$ depth     & 
no. of pixels       & 
\multicolumn{2}{c}{no. of detected galaxies}   & 
\multicolumn{2}{c}{no. of $>5\sigma$ galaxies} \\ 
(sq. degrees) & 
              & 
              & 
$> 5  \sigma$  & 
$> 10 \sigma$  & 
$ z > 1$   & 
$ z > 3$   \\
\hline
1.7    & 5  mJy &  16777  & 1420  & 450 & 1300  & 250 \\
3.3    & 7  mJy &  32884  & 1670  & 480 & 1530  & 250 \\ 
6.8    & 10 mJy &  67111  & 1870  & 500 & 1680  & 250 \\
15.4   & 15 mJy &  151000 & 1890  & 420 & 1740  & 220 \\
61.6   & 30 mJy &  604000 & 1680  & 420 & 1530  & 180 \\
\hline
\end{tabular}
\caption{Estimated number of galaxies detected in test flight and LDB
surveys. Illustrative redshift distributions for galaxies detected
with a $\rm S/N > 5$ are included. The entire LDB flight will be
250~hours allowing for several such surveys.  These
calculations use very conservative estimates of the NEFDs. }
\end{center}
\label{tab:6hour}
\end{table*}

\section{Galactic Plane Surveys}

Over the last decade, sub-mm astronomy has revealed a class of heavily
embedded, pre-stellar cores that are ``invisible'' or faint at FIR
wavelengths, and yet are relatively strong sub-mm
sources \cite{beichman86,awb93,bac99}.   These
low-mass Class~0 and Class~I sources are beginning their main accretion phase
prior to collapse, and hence represent the earliest and most exciting
stages of star-formation.  The mechanism by which these young,
accreting protostars form and evolve is still poorly understood.
{\sl BLAST} will carry out a significant unbiased census of
star-forming cores in a large number of molecular clouds as part of a
sensitive large-area Galactic plane survey.

\begin{figure}[tbh]
\centerline{
\psfig{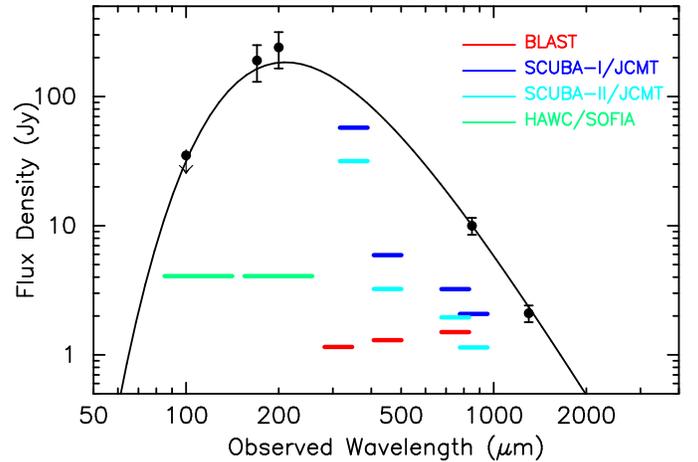}}
\vspace{-0.0in}
\caption{The spectral energy distribution of 
L1544, a pre-stellar embedded core. The horizontal bars represent
the $10\,\sigma$ 1~s sensitivities to extended emission.}
\label{fig:proto}
\end{figure}

By conducting the {\sl BLAST} survey in fields previously observed by
IRAS and ISO, and in the future with SIRTF and SOFIA,
we can determine the relative fraction of Class\,0
and Class\,I sources.  This fraction will indicate the
dynamical timescale for accretion and subsequent collapse.  We can
also trace how this fraction changes with the level of overall
star formation in different molecular cloud complexes, since the
environment of the parent cloud (ionization field, shocks, turbulent
fragmentation) is expected to effect 
the early evolution of the condensing cores.\cite{chan00}

Figure~\ref{fig:proto} illustrates that {\sl BLAST} will have sufficient
sensitivity at 300-450~$\mu$m to
detect the pre-stellar and protostellar clumps within individual molecular
clouds and to measure the
radial density profiles in the extended envelopes of the
protostars.\cite{dwt94} These will discriminate 
between thermally-supported or
magnetically-supported cores.  Molecular lines provide powerful
signatures of infall~\cite{mardones} and the {\sl BLAST} LDB surveys will
provide  a dramatic increase in the
statistical samples of pre-stellar cores for
spectroscopic follow-up at mm-wavelengths.

These data will measure the variations in column densities and
sub-mm spectral indices which reflect variations in the temperatures
(10--30~K) and/or dust emissivity ($\beta=0-2$) within the star forming
envelopes~\cite{awb93}. The combination of longer wavelength surveys ({\it e.g.}
SCUBA/JCMT at 850~$\mu$m and BOLOCAM/CSO at 1100~$\mu$m), and short sub-mm
wavelength data ({\sl BLAST} at 350, 250~$\mu$m) can resolve this ambiguity
since {\sl BLAST} will be more sensitive to temperature variations of
the cold dust (T$< 20$~K).  Having determined the dust temperature we can
then estimate the mass and luminosity distribution of the pre-stellar
clumps. 

Although the origin of the stellar initial mass function is not
well understood, it is probably determined at the earliest stage of
pre-stellar collapse, so it is important to have a large, complete
sample of Class\,0 cores.  In $<$~6 hours {\sl BLAST} will conduct a
$\sim$50~deg$^2$ 300~$\mu$m survey of molecular clouds, including
Orion, Taurus and Ophiuchus, with a 3$\sigma$ sensitivity of
300\,mJy. This is equivalent to a mass sensitivity $\sim
0.05$~M$_{\odot}$.

\section{LDB  Surveys}

The  250~hour LDB flights  will concentrate on 
larger and deeper surveys of Galactic and extragalactic target regions.
The increased observation time and the planned addition of a larger
primary will increase
sensitivities and resolution while lowering confusion limits, thereby  
improving the statistics of the sub-mm source-counts.
In particular, increased counts at faint flux levels will provide the best
discrimination between the possible evolutionary models and  will follow up
on the more extensive SCUBA, SIRTF, AXAF and XMM surveys.  As Table~\ref{tab:sens}
indicates, the improved instrument will detect thousands of high-$z$
galaxies during one of several 50 hour surveys in a single LDB flight. 
Possible LDB extragalactic survey targets include:
(i) for Northern hemisphere flights, the Lockman Hole, and ELAIS N2,
ELAIS N1, the
HDF-North and flanking fields and future LMT 1100~$\mu$m surveys; (ii) for
Southern hemisphere flights,  the MARANO field and the HDF-South.

We will also survey a large ($>25\%$) fraction of
the Galactic plane concentrating on regions undergoing high and low-mass
star formation. 

\section{Instrument}

The design and specifications of {\sl BLAST} are driven by science
goals, availability of
existing instrumentation and the practical limitations of ballooning.
The following sections describe the instrumental
requirements needed to meet the science goals. 

\begin{figure}[t]
\centerline{
\psfig{file=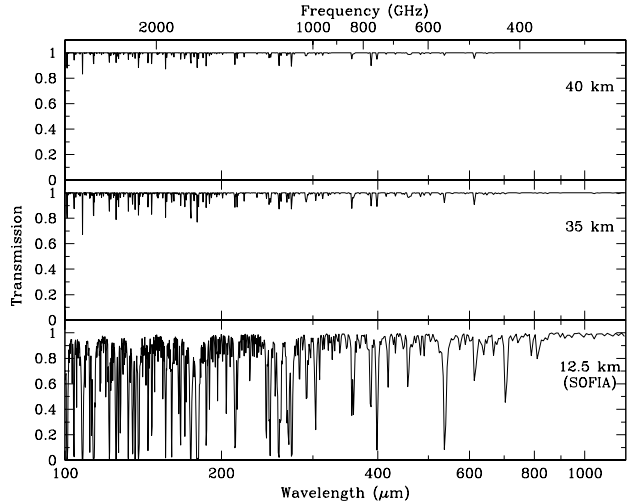,angle=0,width=3.5in}}
\vspace{-.75in}
\caption{Atmospheric transmission at balloon and SOFIA altitudes.} 
\label{fig:atm1}
\end{figure}

	The decision to use an LDB platform for this experiment takes
into consideration a combination of sensitivity, cost, and time-scale.
When comparing this experiment to ground-based or airborne
observations, the clear advantage is greater atmospheric transmission
(Figure \ref{fig:atm1}) at balloon altitudes (35-40~km). 
The high atmospheric emission at lower altitudes (SOFIA and SCUBA)
limits the instrument
sensitivity and, in some of the higher frequency bands, makes the
measurement virtually impossible without the use sub-orbital and orbital
platforms.

\subsection{Optical Design:}

\begin{figure}[tbh]
\centerline{
\hspace{-1.5in}
\psfig{file=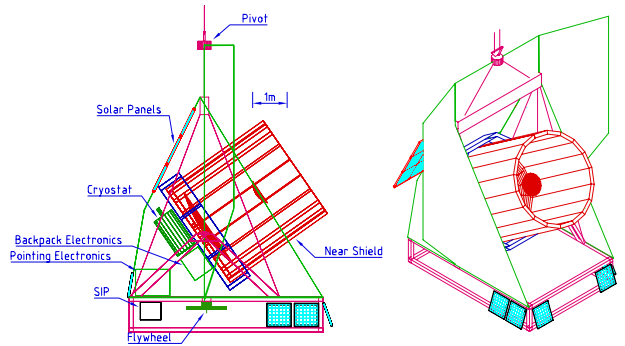,angle=90,width=1.75in}}
\caption{The {\sl BLAST} telescope.} 
\label{fig:tele}
\end{figure}

The {\sl BLAST} gondola is designed to hold a mirror up to
2.5~m in diameter.  The test flight will use 
a 2.0~m spherical mirror currently being fabricated as
part of the {\sl FIRST} development. A schematic of the gondola and
telescope is shown in Figure~\ref{fig:tele}.
The secondary mirror will be 
designed to give diffraction limited performance over a $10'\times10'$
FOV at the Cassegrain focus at $\lambda = 300\, \mu$m.
The estimated antenna efficiency is $\ge 80\%$ and is determined
by a combination of the 10~$\mu$m rms surface roughness of the primary and the
quality of the re-imaging optics.

\begin{table*}[tbh]
\begin{center}
\begin{tabular}{|l|l|l|}\hline
Telescope: &Temperature&300K (230K for North American Flight)\\
\hline
& Used diameter&1.9~m (secondary mirror is pupil stop)\\
\hline
& Emissivity & 0.04\\
\hline
&&\\
\hline
Detectors: & Bolometer optical NEP & $ \rm 3.0\times 10^{-19} \frac{W}{\sqrt{Hz}}$ \\
\hline
& Bolometer quantum efficiency & 0.8 \\
\hline
& Bolometer feed-horn efficiency & 0.7 \\
\hline
& Throughput for each pixel & $\rm A\Omega = \lambda^{2}~~ (2.0f\lambda$ feed-horns)\\
\hline
&& \\
\hline
Bolometers:& Central wavelengths & 250~~350~~500~$\mu$m \\
\hline
& Number of pixels & 149~~88~~~43 \\
\hline
& Beam FWHM & 30~~~41~~~59~arcseconds \\
\hline
& Field of view for each array & 6.5 x 13 arcminutes\\
\hline
& Overall instrument transmission & 30\% \\
\hline
&Filter widths ($\lambda/\Delta \lambda$) & 3 \\
\hline
& Observing efficiency & 90\% \\
\hline
\end{tabular}
\end{center}
\caption{Telescope and Receiver Parameters}
\label{tab:param}
\end{table*}

Radiation from the telescope will enter the cryostat
through a 5--6~cm diameter vacuum window near the Cassegrain focus.  We will
use a telescope focal ratio of f/5 to place the position of
the focus $\simeq$ 20~cm behind the central hole in the primary.
The window will be made from 50~$\mu$m thick polypropylene
that has $<$~0.1\% loss.  Blocking filters at the intermediate cold stages
of 77~K and 20~K will reduce the radiation loading on the
LHe to $<$~10~mW.

\begin{table}[tbh]
\begin{tabular}{|l|l|l|l|}\hline
Band ($\mu$m) &{\bf 250} & {\bf 350}  & {\bf 500}	 \\
\hline
Backgrnd. power~(pW) & 25.6 & 18.3 & 13.5 \\
\hline
Backgrnd.-lim. NEP~$\rm \frac{W}{\sqrt{Hz}}(10^{-17})$ & 20 &
14  & 10\\
\hline
NEFD mJy-$\rm \sqrt s$& 236 & 241 & 239 \\
\hline
$\rm \Delta S (1\sigma , 1 hr)$ (1 deg$^2$) mJy& 38 & 36 & 36 \\
\hline
$\rm \Delta S (1\sigma , 6 hr)$ (1 deg$^2$) mJy& 15.5 & 14.7 & 14.6 \\
\hline
\end{tabular}
\caption{{\sl BLAST} loading, BLIP noise, and Sensitivities}
\label{tab:sens}
\end{table}

The radiation will be re-imaged onto
the detector arrays using a pair of cooled off-axis parabolic mirrors
arranged in a ``Gaussian beam telescope'' configuration.
This configuration makes additional
corrections for aberrations in the main telescope and provides
a flat focal plane with phase centers independent of wavelength
for single mode Gaussian beams.  A cold aperture or
Lyot stop will be located between the two re-imaging mirrors at the
position of an image of the primary mirror to provide additional
sidelobe rejection.  The second parabolic mirror
has a focal length equal to its distance from this aperture to
insure that all of the detectors in the array have illumination
centered on the Lyot stop.
The off-axis angle of the final re-imaging mirror allows us
to place dichroic beam splitters in front of the
focal plane making possible simultaneous measurements by different arrays
at different wavelengths.
For the first flight of {\sl BLAST}, we plan to use detector arrays
similar to those being developed for the SPIRE~\cite{gri00}
instrument on {\sl FIRST}.

\subsection{Detectors.}

The {\sl BLAST} focal plane will consist of
arrays of 149, 88 and 43 detectors at 
250, 350, and 500~$\mu$m respectively. 
The detectors will be silicon nitride micromesh (``spider-web'') bolometric
detectors coupled with $2f\lambda$ feedhorn arrays.\cite{bock98}  The detector and
feedhorn technology is well established and has been tested using
the BOLOCAM instrument in late 1999.  The sensitivity
of the detectors is limited by
photon shot noise from the telescope and atmospheric emission.
Table~\ref{tab:param} gives the telescope and detector parameters.
We estimate a total emissivity for the warm optics of $\simeq 5$\%
dominated by blockage from the secondary mirror and supports.
We estimate the optical efficiency of the cold filters and optics
to be $\epsilon_{\rm{opt}} \ge 0.2$ based on measurements with similar bolometers
coupled to horn arrays at millimeter wavelengths in the BOLOCAM
test dewar.  The estimated detector NEFD's assuming a 10\% bandwidth
are given in  Table \ref{tab:sens}

\section{Conclusion}
{\sl BLAST} will be the first balloon-borne instrument to take advantage
of a new generation of bolometric arrays.  The transparency of the
atmosphere at balloon altitudes will allow it to observe in the
sub-millimeter band where measurements are difficult from the ground.
The combination of state-of-the-art detectors on a sub-orbital observing
platform will give us the opportunity to collect data which will
revolutionize our view of the sub-millimeter sky.  This class of intermediate
missions will provide an integral step in our understanding of this
field in the pre--{\sl FIRST} era.
After an initial test flight in 2002, the instrument will have its first Long
Duration Balloon flight from Antarctica in 2003.

\end{document}